\documentstyle[12pt,a4,psfig,here]{article}
\newcommand{\be}{\begin{equation}}
\newcommand{\ee}{\end{equation}}
\newcommand{\bea}{\begin{eqnarray}}
\newcommand{\eea}{\end{eqnarray}}
\newcommand{\bean}{\begin{eqnarray*}}
\newcommand{\eean}{\end{eqnarray*}}
\newcommand{\ba}{\begin{array}}
\newcommand{\ea}{\end{array}}

\newcommand{\norsl}{\normalsize\sl}
\newcommand{\norsc}{\normalsize\sc}

\begin{document}

\begin{titlepage}
\title{Azimuthal angular dependence of decay lepton\\
       in $e^+ e^- \to t \bar{t}$}

\author{
\norsc  Yuichiro KIYO\\
\norsl  Dept. of Physics, Tohoku University\\
\norsl  Sendai 980-8578, JAPAN\\
\\
\\
\norsc  Jiro KODAIRA and Kazushige MORII\\
\norsl  Dept. of Physics, Hiroshima University\\
\norsl  Higashi-Hiroshima 739-8526, JAPAN\\}

\date{}
\maketitle

\vspace*{1.5cm}

\begin{abstract}
{\normalsize
\noindent
We discuss top quark production and its subsequent decay
for searching new physics at lepton colliders.
The angular dependence of the decay leptons is calculated
including both QCD corrections and anomalous
$\gamma / Z - t \bar{t}$ couplings.
The off-diagonal spin basis for the top and anti-top quarks
is shown to be useful to probe the anomalous couplings.
}

\end{abstract}

\begin{picture}(5,2)(-330,-560)
\put(2.3,-110){TU-600}
\put(2.3,-125){HUPD-0004}
\end{picture}

\vspace{2cm}
\leftline{\hspace{1.2cm}hep-ph/0008065}
\leftline{\hspace{1.2cm}August 2000}

\thispagestyle{empty}

\end{titlepage}
\setcounter{page}{1}
%
 
\section{Introduction}

Since the discovery of the top quark, with a large mass \cite{TEVA},
its properties have been widely discussed to obtain a better
understanding of the electroweak symmetry breaking and to
search for hints of physics beyond the standard model (SM).
It has been known that top quarks decay before
hadronization~\cite{BIGI}.
Therefore there will be sizable angular correlations between
the decay products of the top quark and the spin of the
top quark~\cite{KUHN}.
Based on this observation, it is expected that
we can either test the SM or obtain some signal from new physics
by investigating the angular distributions of the decay products
from polarized top quarks.
Applying the narrow width approximation to the top quarks,
we can discuss the production process and decay process separately.
There are many works on the spin correlations in top quark
production at lepton and hadron colliders~\cite{MP}.
The angular dependence of the decay products from polarized 
top quark has also been discussed~\cite{KUHN,KLY}.

Although it was common to use the helicity basis to decompose
the top quark spin, it has been pointed out by Mahlon, Parke and
Shadmi~\cite{MAHL1} that there is a more optimal decomposition
of the top quark spin depending on the process and the center-
of-mass energy $\sqrt{s}$.
For instance, at a lepton collider, $\sqrt{s}$ 
of which is around several hundred GeV, it was shown~\cite{MAHL1,MAHL2}
that the so-called off-diagonal basis (ODB) is superior
to other bases since top quarks (and/or anti-top quarks) are produced 
in an essentially unique spin configuration.
The QCD one-loop radiative corrections to the
spin correlation in top quark production are also investigated
in ref.\cite{KODA}.
These radiative corrections
induce an anomalous $\gamma/Z$ magnetic moment for the top quarks
and allow for single, real gluon emission.
Therefore, these effects possibly modify the tree level results.
However what was found in ref.\cite{KODA} is that
the effect of the QCD corrections is mainly just the enhancement
of the tree level result and does not change the spin configuration
of produced top quarks (and/or anti-top quarks).
This means that the ODB remains as a good basis even after
including the QCD corrections.

On the other hand, there are also many detailed studies on the
effects of new operators which might come from physics beyond
the SM~\cite{AnoCou1,AnoCou2}.
The fact that the SM is consistent with the data
within the present experimental accuracy tells us that
the size of the effects of new physics is at most comparable to
or smaller than the radiative corrections in the SM.
Therefore, although the QCD correction to the top quark production
is not so large, it should be included to detect these \lq\lq small\rq\rq\ 
signals from possible new physics beyond the SM.

In this article, we investigate the top quark production and 
its subsequent decay at lepton colliders both in the helicity and
off-diagonal basis (ODB) including both the QCD correction and
the assumed anomalous couplings
for the $t \bar{t} - \gamma/Z$ interaction.  
For the angular distribution of the decay products,
the interference between the amplitudes with different spin
configurations plays an important role which disappears
in the production cross section.
We show that the azimuthal angular dependence in the top quark decay 
$t\to b \bar{l} \nu$ is one of the characteristics of the
cross section in the ODB.

The article is organized as follows.
In Section 2, we present the top quark production amplitudes
including both QCD one-loop corrections and the anomalous
couplings for the $t \bar{t} - \gamma/Z$ interaction.
In Section 3, we analyze the angular dependence of the
decay products from the top quark.
Here we compare the results in the ODB with those in the helicity
basis.
Finally Section 4 contains the conclusions. 

\section{Top quark production with QCD corrections and anomalous couplings}

The process we are considering now is, in principle, a very complicated
$e^- e^+ \to 6 \ {\rm particles}$ one.
However, the narrow width approximation
for the top quark, which is valid for $\Gamma_t \ll m_t$ 
(1.02 $\leq \Gamma_t \leq$ 1.56 GeV for 160 
$\leq m_t \leq$ 180 GeV), makes the situation very simple.
Namely, we can separate the physics into the top production
and the decay density matrices~\cite{JEZA}.

Let us first discuss the top quark production (density matrix).
We assume a general form for the $t \bar{t}$-$Z/\gamma$ vertex as, 
\be
  \Gamma^V_\mu =
   g^V \left\{ \gamma_{\mu} \left[ Q_L^V \omega_{-} + Q_R^V \omega_{+}
            \right] + \frac{(t - {\bar{t}})_\mu}{2 m_t} 
     \left[ G_L^V \omega_{-} + G_R^V \omega_{+} \right] \right\}
                 \label{gvp}
\ee
where $t, \bar{t}$ are momenta of the top and top antiquarks, 
$m_t$ is the top mass, $\omega_{\pm} = (1 \pm \gamma_5 )/2$ is
the right/left projection operator,
and $V= Z$ or $\gamma$. 
Here we use the normalization, $g_\gamma = g \sin\theta_W$ and $g_Z=g$
for the coupling constants with $g$ and $\theta_W$ the $SU(2)_L$ 
coupling and Weinberg angle, respectively. The form factor which will
vanish in the zero electron mass limit is neglected.  
For the $e \bar{e}$-$Z/\gamma$ vertex, we use well 
established SM interaction.
At the tree level in the SM, the coupling constants 
$G^{V}_{L,R}$ are zero and
\be
   Q^{\gamma}_{L} = Q^{\gamma}_{R} = Q_t \equiv  \frac{2}{3}, ~~
   Q^Z_L =  Q_L^t \equiv \frac{3-4 \sin\theta_W^2}{6 \cos\theta_W}, ~~
   Q^Z_R =  Q_R^t \equiv - \frac{2 \sin\theta_W^2}{3 \cos\theta_W}.
\label{eq:couplings}
\ee
The combination of form factors  
$G^{\gamma, Z}_R+G^{\gamma, Z}_L \equiv f_2^{\gamma, Z}$ is 
induced even at the one-loop level in the SM.  
Whereas, another combination $G^{\gamma, Z}_R - G^{\gamma, Z}_L
\equiv i f_3^{\gamma, Z}$ which is 
related to a CP violating interaction, called electric 
and weak dipole form factors (EDM and WDM) appears as, at least,
the two-loop order effect.
Thus they are negligibly small and non-zero value of $f_3^{\gamma,Z}$ 
is considered to be a contribution from new physics beyond the SM. 
We presume some non-zero value for $f_3^{\gamma,Z}$
and consider the top production.

To incorporate the QCD one-loop correction into this
analysis, we utilize the fact~\cite{KODA} that
the full one-loop QCD result can be reproduced quite accurately in
the soft gluon approximation (SGA) by choosing an appropriate
cut off $\omega_{\rm max}$ for the soft gluon energy.
There, the formula
$\omega_{\rm max} = \left( \sqrt{s} - 2 m_t \right) / 5$
was suggested.
The difference between the SGA using this $\omega_{\rm max}$ and
the full one-loop correction is smaller than the expected
size of the two-loop corrections.    
In the SGA, all QCD effects can be absorbed into the modified
$t \bar{t}$-$Z/\gamma$ vertex, eq.(\ref{gvp}), using the two
universal functions ${\cal A}$ and ${\cal B}$.
\bea
  Q_{L}^{\gamma} &=& Q_{R}^{\gamma} \, \equiv \, Q^{\gamma}
         = Q_t \, ( 1 + \hat{\alpha}_s \, {\cal A} ) ,\nonumber\\
  Q_L^Z &=& Q_L^t \, ( 1 + \hat{\alpha}_s \, {\cal A} ) + ( Q_L^t - Q_R^t )
           \,   \hat{\alpha}_s \, {\cal B} \ ,\\
  Q_R^Z &=& Q_R^t \, ( 1 + \hat{\alpha}_s \, {\cal A} ) - ( Q_L^t - Q_R^t )
           \,  \hat{\alpha}_s \, {\cal B} \ .\nonumber
\label{eq:qcdvertex}
\eea
and
\be
  G_{L , R}^{\gamma} = Q_t \, \hat{\alpha}_s \, {\cal B} , \quad
   G_{L , R}^Z = \frac{Q_L^t + Q_R^t}{2} \, \hat{\alpha}_s \, {\cal B} \ .
\label{eq:f2}
\ee
where the strong coupling constant is $\hat{\alpha}_s \equiv
\frac{C_2 (R)}{4 \pi} \alpha_s = \frac{C_2 (R)}{(4 \pi )^2} g^2$
with $C_2 (R) = \frac{4}{3}$ for SU(3) of color.
As mentioned before, the one-loop QCD correction does not
contribute to the combination $G^{\gamma, Z}_R - G^{\gamma, Z}_L$.
Since we assume an anomalous coupling to this combination,
eq.(\ref{eq:f2}) is modified to be,
\bea
  G_{L/R}^{\gamma} &=& Q_t \, \hat{\alpha}_s \, {\cal B} \mp \frac{i}{2}
             f_3^{\gamma}\ ,\nonumber\\ 
  G_{L/R}^Z &=& \frac{Q_L^t + Q_R^t}{2} \, \hat{\alpha}_s \, {\cal B} \
               \mp \frac{i}{2} f_3^{Z} .
\label{eq:f3}
\eea
The \lq\lq renormalized\rq\rq\ form factors ${\cal A}$ and ${\cal B}$ read
after multiplying the wave function renormalization factor (we employ
the on-shell renormalization scheme),
\bea
    {\rm Re} {\cal A} &=&      
      \left( \frac{1 + \beta^2}{\beta} \ln
                  \frac{1 + \beta}{1 - \beta} - 2 \right)
              \ln \frac{4 \omega_{\rm max}^2}{m_t^2} - 4 + 
           \frac{2 + 3 \beta^2}{\beta}
              \ln \frac{1 + \beta}{1 - \beta}  \nonumber\\
       &+&  \frac{1 + \beta^2}{\beta}
            \left\{ \ln \frac{1 - \beta}{1 + \beta}
               \left( 3 \ln \frac{2 \beta}{1 +  \beta}
         +  \ln \frac{2 \beta}{1 - \beta} \right) 
         + 4 {\rm Li}_2 \left( \frac{1 - \beta}{1 + \beta} \right)
          + \frac{1}{3} \pi^2 \right\} \ ,\label{afactor}\\
    {\rm Im} {\cal A} &=& \pi\,\left( 
         -3 \beta + \frac{1 + \beta^2}{\beta} \ln
             \frac{4 \beta^2}{1 - \beta^2} - \frac{1 + \beta^2}{\beta}
                 \ln \frac{\lambda^2}{m_t^2}\right)  \ ,\nonumber\\
    {\rm Re} {\cal B} &=& 
          \frac{1 - \beta^2}{\beta} \ln \frac{1 + \beta}{1 -
                      \beta}\ ,\label{bfactor}\\
    {\rm Im} {\cal B} &=& \pi \, \frac{\beta^2 -1}{\beta}\ ,\nonumber
\eea
where $\beta$ is the speed of the produced top (anti-top)
quark.
In the form factor ${\cal A}$ eq.(\ref{afactor}), we have already
took into account the contribution from the real gluon emission.
Therefore, there is no infrared singularity in the real part
and instead there appears
the soft gluon cut-off $\omega_{\rm max}$.
We have introduced an infinitesimal mass $\lambda$ for the gluon to avoid
the infrared singularity which remains in the imaginary part of ${\cal A}$. 
However it will be shown that the imaginary part of ${\cal A}$ does not
contribute to any observable within our approximation 
which keep only the linear terms in $\alpha_s$ and $f_3^{\gamma, Z}$.

\begin{figure}[H]
\begin{center}
\begin{tabular}{cc}
\leavevmode\psfig{file=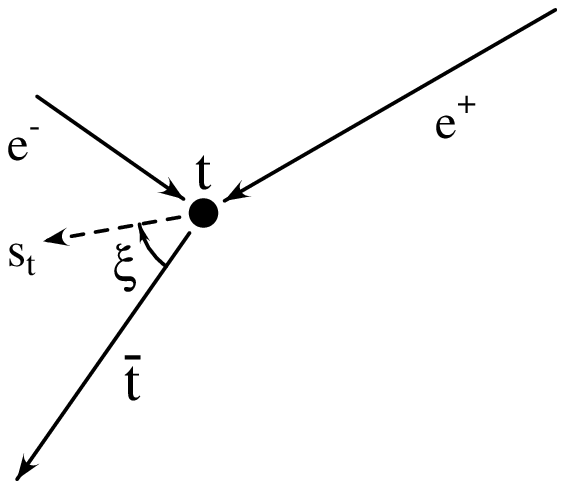,width=5cm} &
\leavevmode\psfig{file=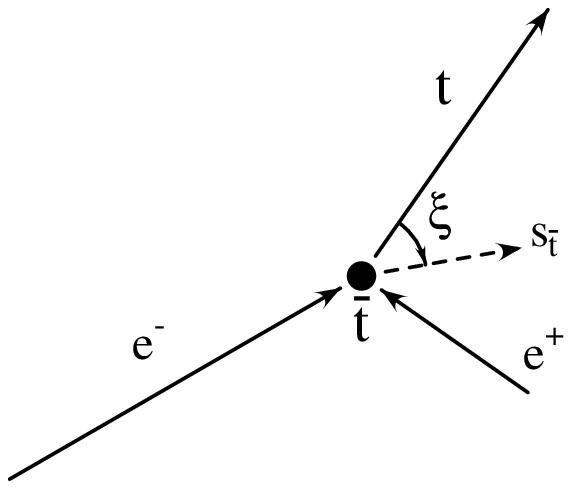,width=5.5cm} 
\end{tabular}
\caption{The generic spin basis for the top (anti-top)
quark in its rest frame.  ${\bf s}_t$ (${\bf s}_{\bar{t}}$) is the top
(anti-top) spin axis.}
\label{fig:spin}
\end{center}
\end{figure}

Before presenting the production amplitudes,
let us define the spin basis according to ref.~\cite{MAHL1}.
In this paper, we consider the case in which the spin of the 
top quark and anti-top quark in the production plane.
The spins of the top and anti-top quarks are parameterized by $\xi$
as given in Fig.\ref{fig:spin}.
The top quark spin is decomposed along the direction ${\bf s}_t$
in the rest frame of the top quark which makes an angle
$\xi$ with the anti-top quark momentum in the clockwise direction.   
Similarly, the anti-top quark spin states are defined in the anti-top
rest frame along the direction ${\bf s}_{\bar{t}}$ having the same
angle $\xi$ from the direction of the top quark momentum.
The state
$t_{\uparrow}\,\bar{t}_{\uparrow}\,(t_{\downarrow}\,\bar{t}_{\downarrow})$
refers to a top with spin in the $+ {\bf s}_t \,(- {\bf s}_t )$
direction in the top rest frame and an anti-top
with spin $+ {\bf s}_{\bar{t}} \,(- {\bf s}_{\bar{t}} )$
in the anti-top rest frame.
Note that the value $\cos \xi = - 1$ corresponds to the helicity state.
For the initial leptons, we use 
the helicity basis with the notation $e^{+}_{R,L}$ and $e^-_{R,L}$,
where the subscripts $R, L$ denote the helicities of the particles.
Now, the production amplitudes of top quark pairs
in $e^-_L e^+_R$ annihilation
turns out to be written in the following general forms
in the zero momentum frame (ZMF),
\bea
M(e_L^-e_R^+\rightarrow t_\uparrow \bar{t}_\uparrow,
                        t_\downarrow \bar{t}_\downarrow)
  &=&  \mp 4\pi \alpha \left[ \left( {A}_{LR} - {C}_{LR} \right)
 \cos \xi - {B}_{LR}
           \sin \xi \pm i {E}_{LR} \right] \ ,\label{eq:amp_1}\\
M(e_L^-e_R^+\rightarrow t_\uparrow \bar{t}_\downarrow, 
                        t_\downarrow \bar{t}_\uparrow)
  &=& \ \ 4\pi \alpha \left[ \left( {A}_{LR} - {C}_{LR} \right)
 \sin\xi +  {B}_{LR} \cos\xi
           \pm {D}_{LR} \right] \ , \label{eq:amp_2}
\eea
employing an appropriate phase convention for
spinors~\cite{MAHL2,AnoCou1}.
$\alpha$ is the QED structure constant.
The coefficients ${A}_{LR}$ , ${B}_{LR}$
${D}_{LR}$ and ${E}_{LR}$ are given by,
\bea
  {A}_{LR} &=& \frac{1}{2}\, 
  \Bigl[ ( {f}_{LL} + {f}_{LR} ) \sqrt{1 - \beta^2}
            \sin \theta \Bigr] \ , \nonumber\\
  {B}_{LR} &=& \frac{1}{2}\,
       \Bigl[ {f}_{LL} ( \cos \theta + \beta )
             + {f}_{LR} ( \cos \theta - \beta ) \Bigr] \ ,\nonumber\\
  {C}_{LR} &=& \frac{1}{2} \,
              (h_{LL} + h_{LR}) \frac{\beta^2 \sin \theta}
                         {\sqrt{1 - \beta^2}} \ , \label{parame1}\\
  {D}_{LR} &=& \frac{1}{2} \,
            \Bigl[ {f}_{LL} ( 1 + \beta \cos \theta )
             + {f}_{LR} ( 1 - \beta \cos \theta ) \Bigr] \  ,\nonumber\\
  {E}_{LR} &=& \frac{- i}{2} (h_{LL} - h_{LR}) \frac{\beta \sin \theta}
            {\sqrt{1-\beta^2}} \ ,\nonumber
\eea
with
\bea
  {f}_{IJ} &=& - Q^{\gamma} + \frac{Q^e_I \, Q^Z_J}{\sin^2 \theta_W}
              \frac{s}{s - M_Z^2}\ ,\nonumber\\
       h_{IJ} &=&  - G^{\gamma}_J + \frac{Q^e_I \, G^Z_J }
             {\sin^2\theta_W} \frac{s}{s - M_Z^2} \ , \label{parame2}
\eea
where the angle $\theta$ is the scattering angle
of the top quark with respect to the electron in the ZMF.
$M_Z$ is the Z-boson mass and we neglect the 
$Z$ width since it is negligible in the region
of center-of-mass energy $\sqrt{s}$ far above the production
threshold for top quarks.
$I,J \in (L,R)$ and $Q^e_I$ is the electron coupling to the
Z boson given by,
\[   Q^e_L = \frac{2 \sin^2 \theta_W - 1}{2 \cos \theta_W} \quad ,\quad 
     Q^e_R = \frac{ \sin^2 \theta_W}{ \cos \theta_W} \ .\]
In eq.(\ref{parame1}), ${C}_{LR}$ is proportional to 
$f_2^{\gamma,Z}$ and the contribution from CP violating form 
factors $f_3^{\gamma,Z}$ enter through ${E}_{LR}$ 
only. The contributions from $f_2^{\gamma, Z}$, WDM and EDM 
form factors $f_3^{\gamma,Z}$, are enhanced when $\beta$ becomes 
large, and become zero for $\beta \to 0$. 
The amplitudes for $e^-_R e^+_L$ can be obtained by interchanging 
$R$ and $L$ as well as $\uparrow$ and $\downarrow$ in 
Eqs.(\ref{eq:amp_1})--(\ref{parame2}). 

At the tree level in the SM (${\cal A} = {\cal B} =  f_3 = 0$),
the coefficients ${C}_{LR}$ and ${E}_{LR}$ are zero
and other coefficients become real in eqs.(\ref{eq:amp_1},\ref{eq:amp_2}).
The ODB for the process 
$e^-_L e^+_R\rightarrow t \bar{t}$ is defined by, 
\[ \tan\xi = \left. \frac{{A}_{LR}}{{B}_{LR}} \right|_{\rm tree} \ , \]
which makes the like-spin configurations 
$t_{\uparrow}\bar{t}_{\uparrow}$ and 
$t_{\downarrow}\bar{t}_{\downarrow}$ be zero.
The up-down ($t_{\uparrow}\bar{t}_{\downarrow}$) configuration dominates the 
cross sections in the ODB whereas the down-up 
($t_{\downarrow}\bar{t}_{\uparrow}$) is numerically negligible,
less than $1\%$ of the total cross section. 
 
The problem now is how to detect the anomalous coupling in
the top quark events.
It is easily understood that the effects of the anomalous coupling
on the top quark production cross sections should be small and
undetectable because (1) the anomalous coupling is assumed to be comparable
to or smaller than the QCD correction in size and we already know
the QCD correction itself to be very small and (2) the interference
terms disappear in the production cross sections.
Therefore we consider the angular distribution of top decay products
which depends on the interferences between various amplitudes.

\section{Decay distribution with anomalous coupling}

In the decay process, we assume V-A interaction of the SM in 
$t$-$b$-$W$ vertex. We employ the semi-leptonic decay, 
$t \to b W \to b \bar{l}\nu$ for simplicity. 
Neglecting the mass of the final state fermions,
the decay amplitude $D_{s_{t}}$ 
(for $t_{s_t} \to b \bar{l} \nu$) is known to be given by 
\be
   D_{\uparrow}  = \frac{2 g^2 V_{tb} \sqrt{ b \cdot \nu ~m_t E_{\bar{l}}}
      }{2 \nu \cdot \bar{l} - M_W^2 + i M_W \Gamma_{W}} 
        \cos\frac{\theta_{\bar{l}}}{2} \ ,\ 
  D_{\downarrow}  = \frac{ 2 g^2 V_{tb} \sqrt{ b \cdot \nu ~m_t E_{\bar{l}}}
      }{ 2 \nu \cdot \bar{l} - M_W^2 + i M_W \Gamma_{W} } 
      \sin\frac{ \theta_{\bar{l}}}{2} e^{ -i \phi_{\bar{l}} } \ ,\label{eq:D2}
\ee 
where the names of final particles are used as substitute
for their momenta.
$M_W (\Gamma_W )$ and $V_{tb}$ are the mass (width) of the
W boson and the Cabbibo--Kobayashi--Maskawa matrix.
\begin{figure}[htbp]
\begin{center}
\leavevmode\psfig{file=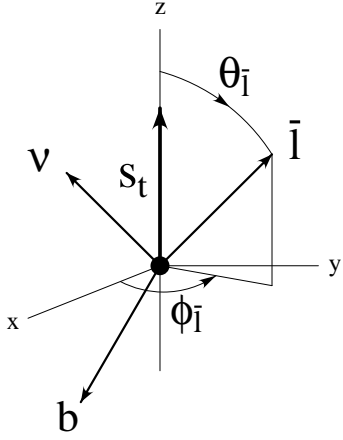,width=4.5cm}
\vspace{-0.6cm}
\caption{The definition of polar and azimuthal angles.}
\label{fig:angle}
\end{center}
\end{figure}
The polar and azimuthal angles of the $\bar{l}$ momentum
($\theta_{\bar{l}}, \phi_{\bar{l}}$) are defined 
in the top quark rest frame, 
in which $z$-axis coincides with the chosen spin axis $s_t$
and the $x-z$ plane is the production plane, Fig.\ref{fig:angle}.
We have a similar expression $\bar{D}_{\uparrow\downarrow}$
also for the anti-top quark decay.

Now, the differential cross-section for the process $e^- e^+ \to 
t \bar{t}$ followed by the decays $t \to X_t\,,\,\bar{t} \to \bar{X}_t$
is described in terms of the production
and decay density matrices
$\rho_{s_t \bar{s}_t , s'_t \bar{s}'_t}$\, ,\,
$\tau_{s_t s'_t}$ and $\bar{\tau}_{\bar{s}_t \bar{s}'_t}$ as,
\be
  d \sigma \left( e^- e^+ \to t\bar{t} 
               \to X_t \bar{X}_t \right)
     \propto \sum_{s_t, \bar{s}_t,s'_t,\bar{s}'_t}
           \rho_{s_t \bar{s}_t , s'_t \bar{s}'_t}
          \tau_{s_t s'_t} \bar{\tau}_{\bar{s}_t \bar{s}'_t} d L \ ,
         \label{prodecayamp}
\ee
where $d L$ is the phase space of the final particles
and the density matrices can be obtained from
eqs.(\ref{eq:amp_1}), (\ref{eq:amp_2}) and (\ref{eq:D2})~\cite{JEZA}. 
\bea
 \rho_{s_t \bar{s}_t , s'_t \bar{s}'_t}
          &=& M_{ s_t \bar{s}_t} M^{\ast}_{s'_t \bar{s}'_t} \nonumber\ ,\\
  \tau_{s_t s'_t} &=&
     D_{s_t} D^\ast_{s'_t} \propto   \left( \ba{cc}
        1+\cos\theta_{\bar{l}} & \sin\theta_{\bar{l}} \, e^{i\phi_{\bar{l}}}\\
       \sin\theta_{\bar{l}}\, e^{-i\phi_{\bar{l}}}& 1-\cos\theta_{\bar{l}} \\
    \ea \right)_{s_t s'_t} \label{decaydm}\ .
\eea
$\bar{\tau}_{\bar{s}_t \bar{s}'_t}$ is also given by the similar
expression.
When we calculated the production density matrix,
we have kept only terms which are linear in $\alpha_s$ and $f_3^{\gamma,Z}$
for the consistency of our approximation.
Within this approximation, the factor $1 + \hat{\alpha}_s A$
can be effectively factorized from the amplitudes as a multiplicative factor.
Therefore its imaginary part which has the infrared divergence
does not contribute to the production density matrix.

\begin{figure}[H]
\begin{center}
\begin{tabular}{cc}
\leavevmode\psfig{file=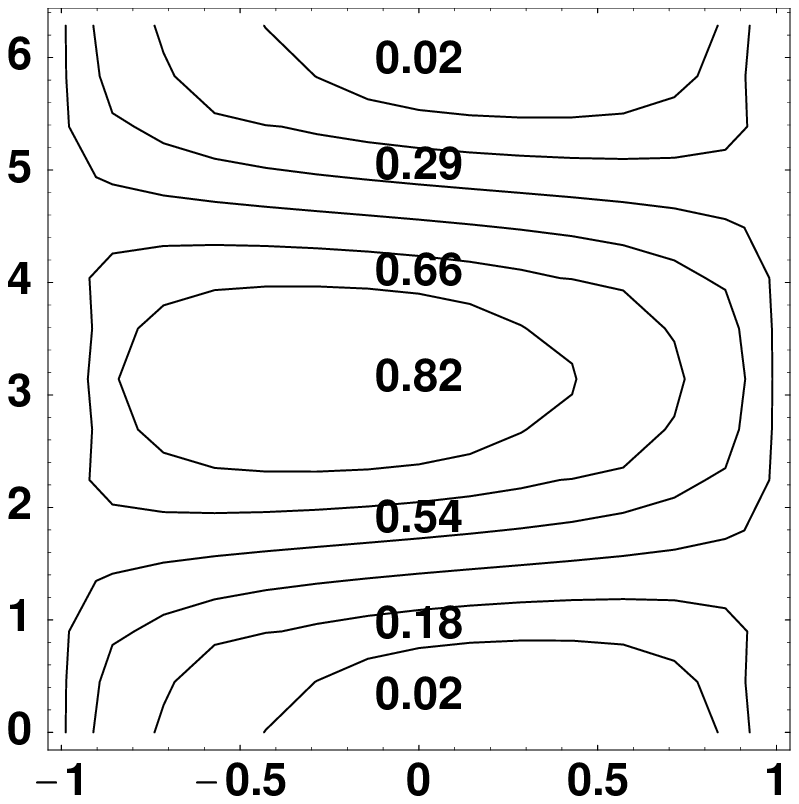,width=6.5cm}  & 
\leavevmode\psfig{file=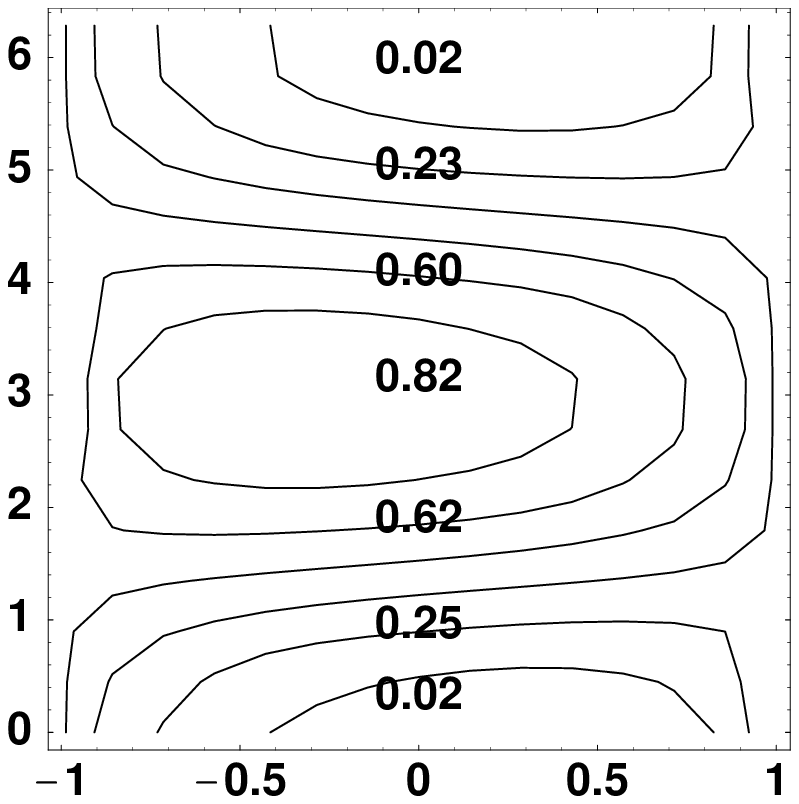,width=6.5cm}
\end{tabular}
\caption{The double differential cross section
$d \sigma / d \cos \theta_{\bar{l}} d \phi_{\bar{l}}$
in the helicity basis.
The left (right) figure correspond to the
cross section without (with) the anomalous $f_3^{\gamma,Z}$
coupling. Vertical and horizontal axes
correspond to the azimuthal $\phi_{\bar{l}}$ and 
the polar angle $\cos\theta_{\bar{l}}$, respectively.}
\label{helcorr}
\end{center}
\end{figure}

\begin{figure}[H]
\begin{center}
\begin{tabular}{cc}
\leavevmode\psfig{file=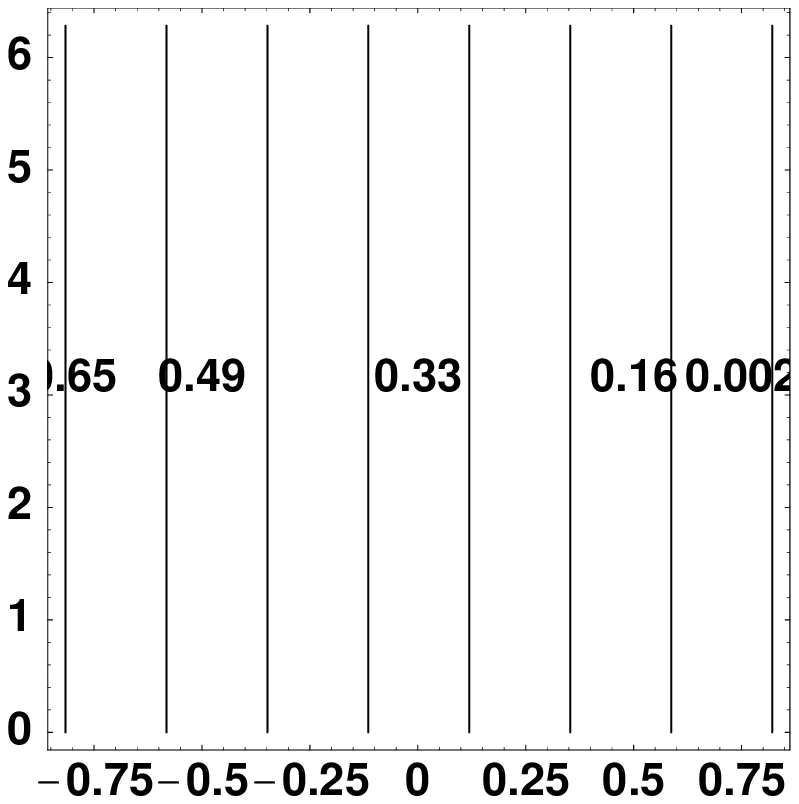,width=6.5cm}  &
\leavevmode\psfig{file=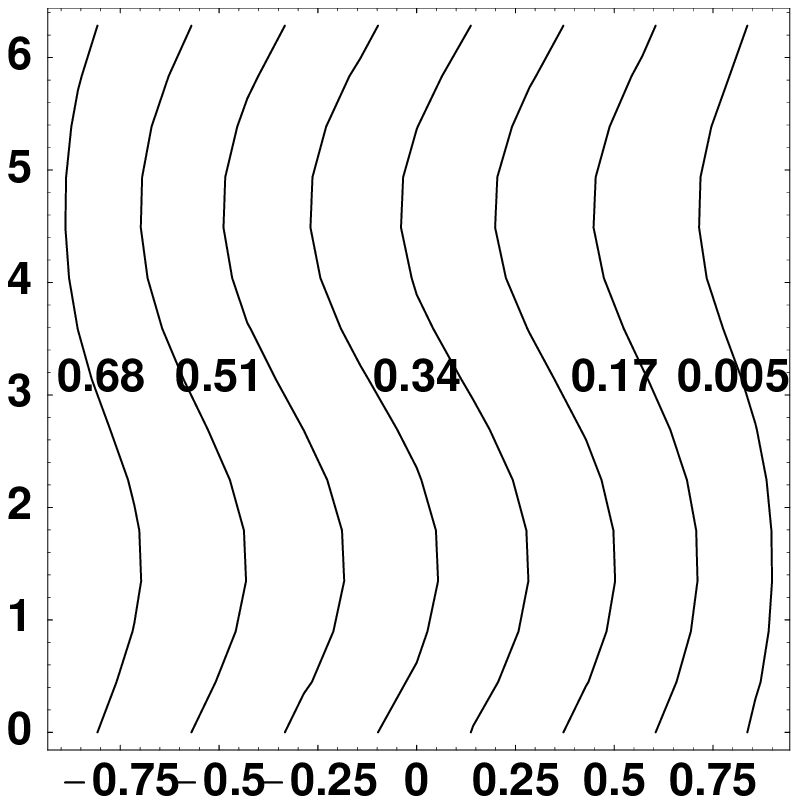,width=6.5cm}
\end{tabular}
\caption{The double differential cross section
in the off-diagonal basis.
The left (right) figure correspond to the
cross section without (with) the anomalous $f_3^{\gamma,Z}$
coupling.
The axes are the same as in Fig.\ref{helcorr} }
\label{ODBcorr}
\end{center}
\end{figure}

Here we take an advantage of the freedom for the choice of the 
spin basis to detect the effect of the anomalous
couplings~\cite{MAHL2,AnoCou1}.
Note that the differential cross section itself is (should be)
independent of the choice of the spin basis.
However, the \lq\lq choice of the variables\rq\rq\  can depend on
the spin basis.
We have calculated the angular distribution of $\bar{l}$
in the top quark decay after integrating out other variables,
\[ \frac{d \sigma \left( e_L^- e_R^+ \to t \bar{t} \to 
                        \bar{l} + X  \right)}
       {d \cos \theta d \cos \theta_{\bar{l}} d \phi_{\bar{l}}} \ .\]
All input masses and coupling constants used in the numerical
computations are the central values as reported in the 1998
Review of Particle Properties~\cite{PDG}.
We plot the $\theta_{\bar{l}} - \phi_{\bar{l}}$ correlations
both in the helicity (Fig.\ref{helcorr})
and the off-diagonal basis (Fig.\ref{ODBcorr}).
We set $\sqrt{s} = 400$ GeV and assume $f_3^{\gamma,Z}= 0.2$
just for an illustration.
The both figures are for $\cos \theta = 0$.
However the pattern of the correlation is essentially the same for
all scattering angles.
One can see that it is very hard to identify the effects of the
anomalous couplings in Fig.~\ref{helcorr}. 
This situation changes drastically if we take the
ODB (Fig.~\ref{ODBcorr}).  
As the SM produces almost no
azimuthal angular dependence in this basis (although the QCD corrections
produce some dependence, it is numerically negligible),
we can recognize the effect 
of the anomalous coupling as a deviation from the flat distribution.
The above results are easily understood if one notices
that the azimuthal angular dependence is caused by
interference effects in a given spin basis.
From eq.(\ref{decaydm}), the azimuthal angular dependence
comes from the off diagonal $\uparrow\downarrow$ or
$\downarrow\uparrow$ element of the decay density matrix.
On the other hand, the production amplitudes, therefore the
density matrix, take the following characteristic
forms (see eqs.(\ref{eq:amp_1},\ref{eq:amp_2})) in the ODB,
\bean
 M_{\uparrow\uparrow} &\sim& M_{\downarrow\downarrow} \sim
                {E}_{LR}\ ,\\
 M_{\uparrow\downarrow} &\sim& {\rm finite}\ , \ 
            M_{\downarrow\uparrow} \sim 0 \ ,
\eean
except small contributions from the QCD correction.
This means that the azimuthal angular dependence
receives significant contributions only from the elements,
\[ \rho_{\uparrow\downarrow , \downarrow\downarrow}\ ,\quad 
   \rho_{\downarrow\downarrow , \uparrow\downarrow} \ ,\]
of the production density matrix and it linearly depends on 
${E}_{LR}$, namely $f_3^{\gamma, Z}$.
The $\phi_{\bar{l}}$ dependence is controlled by the value of
$f_3^{\gamma, Z}$. 
At the tree level, it has a simple form,
\[ {\rm Re}\, f_3^{\gamma, Z} \sin\,\phi_{\bar{l}} - 
            {\rm Im} \,f_3^{\gamma, Z}\cos\,\phi_{\bar{l}} \ .\]

\begin{figure}[H]
\begin{center}
\leavevmode\psfig{file=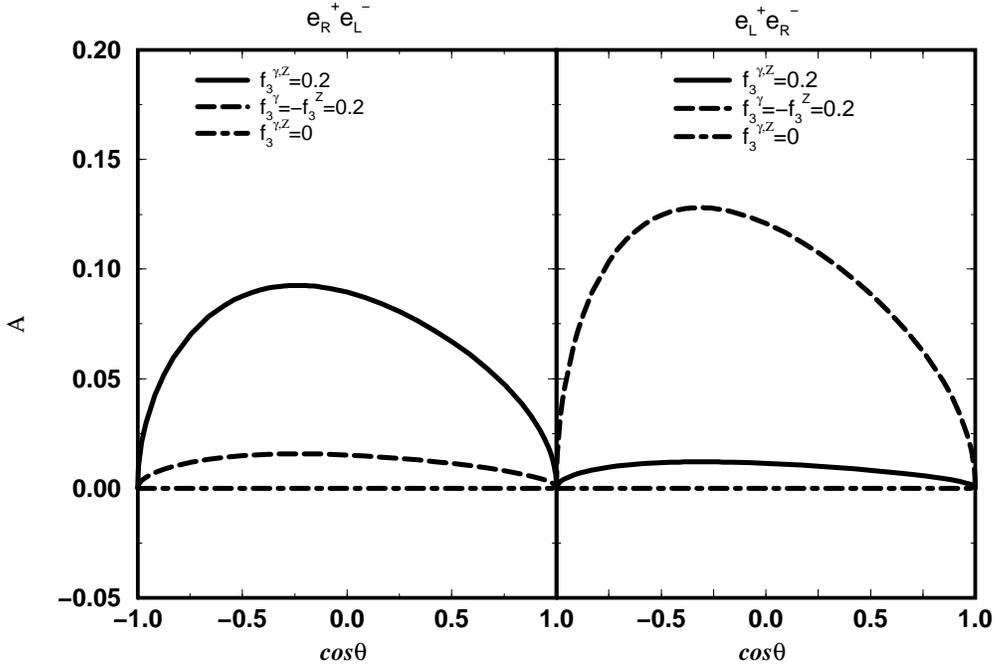,width=15cm,angle=-90}
\vspace{-2cm}
\caption{Azimuthal asymmetry as a function of $\cos \theta$ 
in the off-diagonal basis.} 
\label{fig5} 
\end{center}
\end{figure}

\begin{figure}[H]
\begin{center}
\leavevmode\psfig{file=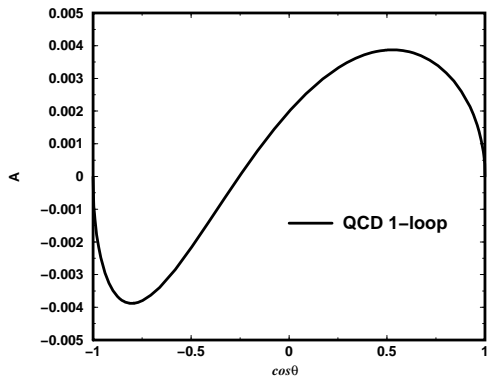,width=7cm}
\vspace{-0.6cm}
\caption{Azimuthal asymmetry induced by the QCD correction
in the off-diagonal basis for the $e^-_L e^+_R$ 
annihilation.} 
\label{QCDasym} 
\end{center}
\end{figure}

In order to show the effect of the $f_3^{\gamma,Z}$ 
more clearly, we partially integrate the cross section over the 
azimuthal angle and define the azimuthal asymmetry. 
Let $\sigma^{1,2}$ denote the partially integrated cross-sections 
over the azimuthal angle,
\bean
 \sigma^1 (\theta) &=& \int_{0}^{\pi} d \phi_{\bar{l}} 
       \left( \frac{d \sigma}{d \cos\theta d \phi_{\bar{l}}} \right)\ ,\\
 \sigma^{2} (\theta) & =& \int_{\pi}^{2\pi} d \phi_{\bar{l}} 
       \left( \frac{d \sigma}{d \cos\theta d \phi_{\bar{l}}}\right)\ ,
\eean
where other variables have been integrated out already.
We define the azimuthal asymmetry in order to pull out the 
effect of anomalous interactions,
\[   {\cal A} (\theta)  = 
  \frac{\sigma^{2} (\theta ) -\sigma^{1} (\theta )}
    {\sigma^{2} (\theta ) + \sigma^{1}(\theta )} . \]
We plot the asymmetry as a function of $\cos\theta$ in 
Fig.\ref{fig5} at 
$\sqrt{s} = 400$ GeV for the $e^-_L e^+_R$ and $e^+_L e^-_R$ 
annihilation.
We have assumed two cases for the anomalous couplings,
${\rm Re}\, f_3^{\gamma} = {\rm Re}\, f_3^Z = 0.2$ and
${\rm Re}\, f_3^{\gamma} = - {\rm Re}\, f_3^Z = 0.2$.
\noindent
In this figure, the dot-dashed line comes from 
the SM (with QCD corrections) and, the solid (dashed) line
corresponds to the case ${\rm Re}\, f_3^{\gamma} = {\rm Re}\, f_3^Z = 0.2$ 
(${\rm Re}\, f_3^{\gamma} = - {\rm Re}\, f_3^Z = 0.2$).
At the SM tree level, the asymmetry is exactly zero and
the QCD radiative corrections induce a numerically negligible
asymmetry as shown in Fig.\ref{QCDasym}.
As explained before, the asymmetry linearly depends on the absolute value
of $f_3^{\gamma,Z}$ and also their sign.
In the case of $e^-_L e^+_R $, the effect 
of the anomalous interactions $f_3^\gamma $ and $f_3^Z$ 
are additive and have a larger asymmetry 
when their signs are the same.
But when their signs are opposite, 
these effects become subtractive and lead to a smaller asymmetry.  
This feature changes in the case of $e^+_L e^-_R$.  
In the off-diagonal basis, the anomalous couplings 
produce the asymmetry of the order 10\% for the values of the
anomalous couplings we have chosen.
In the helicity basis, however, the deviation from the SM 
is only around 1.5\% since there exists some amount of asymmetry 
already in the SM. 
If we take the asymmetry by defining $\sigma^{1,2}$ as,
\bean
 \sigma^1 (\theta) &=& \left( \int_{0}^{\frac{\pi}{2}}
          +  \int_{\frac{3\pi}{2}}^{2\pi} \right)\, d \phi_{\bar{l}} 
       \left( \frac{d \sigma}{d \cos\theta d \phi_{\bar{l}}} \right)\ ,\\
 \sigma^{2} (\theta) & =& \int_{\frac{\pi}{2}}^{\frac{3\pi}{2}} d \phi_{\bar{l}} 
       \left( \frac{d \sigma}{d \cos\theta d \phi_{\bar{l}}}\right)\ ,
\eean
we can obtain information for ${\rm Im}\, f_3^{\gamma, Z}$.

\vspace{-0.5cm}

\section{Conclusion}
We have studied the top quark pair production and subsequent decays 
at lepton colliders.
For a realistic next lepton collider, let us say 
$\beta \sim 0.5$, the off-diagonal basis is
considered to be a good choice since the contribution from
some spin states is zero or negligible even after
including the QCD corrections and 
this small interference makes the correlations between decay products
and the top spin very strong.
Using this advantage, we analyzed the angular dependence
of the decay product of the top quark including
both the QCD corrections and the anomalous couplings. 
We have shown that the asymmetry amount to the order of 10\%
in the off-diagonal basis with chosen parameters which
may be detectable.

Although we have considered the anomalous couplings only
for the production process,
the inclusion of new effects to the decay process
and more detailed and/or realistic phenomenological analyses for various
choices of the new interactions
are straightforward exercises.

\section*{Acknowledgment}

We would like to thank S. Parke for valuable
suggestions.
Y. K. was supported by the Japan Society for the Promotion of Science.


\end{document}